\documentclass[aps,prb,a4paper,preprint,showpacs]{revtex4}
\usepackage{amsmath}
\usepackage{amsfonts}
\usepackage{amssymb}
\usepackage{slashbox}
\usepackage{graphicx}
\usepackage{amsbsy}

\begin{document}
\title{Diffraction induced Spin Pumping in Normal-Metal/Multiferroic-Helimagnet/Ferromagnet
Heterostructures}
\author{Rui Zhu\renewcommand{\thefootnote}{*}\footnote{Corresponding author.
Electronic address:
rzhu@scut.edu.cn}}
\address{Department of Physics, South China University of Technology,
Guangzhou 510641, People's Republic of China }

\begin{abstract}

Generally the adiabatic quantum pumping phenomenon can be interpreted by the surface
integral of the Berry curvature inside the cyclic loop. Spin angular momentum flow without charge current
can be pumped out by magnetization precession in ferromagnet-based structures. When an
electron is scattered by a
helimagnet, spin-dependent diffraction occurs due to the spatial modulation of the spiral.
In this work, we consider the charge and spin flow driven by magnetization precession
in normal-metal/multiferroic-helimagnet/ferromagnet heterostructures. The pumping behavior is governed by the diffracted states. Gauge dependence in the pumped current was encountered, which does not occur in the static transport properties or pumping behaviors in other systems.

\end{abstract}


\maketitle

\narrowtext

\section{Introduction}

A dc charge and spin current can be generated by cyclic variation of system parameters\cite{Ref1, Ref2, Ref14}. In the adiabatic condition, i.e., the escape rate of the particle is much smaller than the speed of the parameter variation, the transport procedure can be viewed as the accumulated effect of static tunneling with the time-dependent parameter frozen at a certain value\cite{Ref5}. The pumped current can be evaluated from the surface integral of the scattering Berry curvature\cite{Ref3, Ref4} or by Taylor's expansion of the instant scattering matrix at the equilibrium parameter values\cite{Ref5, Ref31}. For large pumping frequencies, Floquet scattering theory\cite{Ref6, Ref32, Ref33} and the Green's function technique\cite{Ref9, Ref11} are developed covering both the adiabatic and nonadiabatic situations. To this date, the pumping behavior in almost all novel quantum states has been studied, from the quantum Hall liquids\cite{Ref34} to superconductors\cite{Ref36, Ref37}, from graphene\cite{Ref17, Ref18, Ref19} to topological insulators\cite{Ref35, Ref20}, and etc. Along with theoretical development, quantum charge and spin pumps were realized in various nanoscale transport systems such as the quantum dot\cite{Ref14, Ref15, Ref25}, superconductors\cite{Ref16, Ref36, Ref37}, spin pumping driven by precessing ferromagnet\cite{Ref26}, and etc. The quantum pumping process acts as a platform to display quantum novelties of different quantum states and structures.

As a bifurcation of quantum pumping, research on spin pumping with its counterpart spin transfer torque prospers in its own direction due to its promising spintronic applications\cite{Ref22}. Despite its practical significance, one naturally wonders its role in revealing unknown properties of novel magnetic structures such as domain walls\cite{Ref38}, multiferroic helimagnets\cite{Ref39, Ref40}, and skyrmion lattices\cite{Ref41} etc. Usually spin pumping was investigated in ferromagnet-based multi-layer structures driven by magnetization precession in which spin polarization is uniform in space. In these structures spin momentum flow can be generated with vanishing net charge current. Tserkovnyak et al. considered\cite{Ref45} the time-dependent magnetic order parameter driven quantum pumping properties in helimagnets and analyzed the evolution of the magnetic spiral, in which the broken symmetry is different from the precessing ferromagnet driven helimagnet heterostructures. Spatially nonuniform spin structure induces diffracted transmission in spiral helimagnets\cite{Ref42, Ref43}. It can be predicted that the skyrmion lattice should display sophisticated transmission spectra due to its rich Fourier components of the spin vortex in space. Little in literature addressed the quantum pumping properties featured by diffraction. The topic interests us a lot. So in this work, we consider the spin pumping properties in normal-metal/multiferroic-helimagnet/ferromagnet heterostructures and investigate the physical properties induced by diffracted transmission or what can be coined as spatial nonadiabaticity and the approach can be extended to the skyrmion-lattice-based heterostructures.

\section{Theoretical formulation}

We consider the normal-metal/multiferroic-helimagnet/ferromagnet (NM/MF/FM) triple-layer heterostructure depicted in Fig. 1. The Hamiltonians in different layers can be formulated as:
\begin{equation}
\begin{array}{c}
 \begin{array}{*{20}c}
   {H_{\texttt{NM}}  =  - \frac{{\hbar ^2 }}{{2m_e }}\nabla ^2 ,} & {z < 0,}  \\
\end{array} \\
 \begin{array}{*{20}c}
   {H_{\texttt{MF}}  =  - \frac{{\hbar ^2 }}{{2m^* }}\nabla ^2  + J{\bf{n}}_{\bf{r}}  \cdot {\bf{\sigma }} + V_0 ,} & {0 \le z \le d,}  \\
\end{array} \\
 \begin{array}{*{20}c}
   {H_{\texttt{FM}}  =  - \frac{{\hbar ^2 }}{{2m_e }}\nabla ^2  - \Delta {\bf{m}} \cdot {\bf{\sigma }},} & {z > d,}  \\
\end{array} \\
 \end{array}
 \label{eq1}
 \end{equation}
where $d$ is the thickness of the MF
layer. $m_{e}$ and $m^{*}$ are the free and multiferroic oxides' effective electron masses respectively. ${\bf{\sigma}}$
is the Pauli vector. ${\bf{m}} = \left[ {\sin \theta \cos \phi
,\sin\theta \sin \phi ,\cos \theta } \right]$ is the magnetization unit vector in the FM layer with respect to the
[100] crystallographic direction. $\Delta$ is the half width of the
Zeeman splitting in the FM electrode. $J{\bf{n}}_{\bf{r}} $ is the space-dependent
exchange field following the helicity of the MF spiral with ${\bf{n}}_{\bf{r}}  =  \left[ {\sin \theta _r ,0,\cos \theta _r } \right]$,
 $\theta _r = \bar q_m  \cdot {\bf{r}}$, and $\bar q_m  = \left[
{\bar q,0,0} \right]$. From Eq. (\ref{eq1}) it can be seen that the exchange coupling between the electron and
the localized noncollinear magnetic moments within the barrier acts as a nonhomogenous magnetic field. Therefore, spin-dependent diffraction of transmission can be foreseen in the situation.

We consider an ultrathin film of MF-helimagnet with thickness $d=2$ nm, which can be approximated by a Dirac-delta
function. The MF barrier reduces to a plane
barrier. Its Hamiltonian can be rewritten as
\begin{equation}
H_{MF}  =  - \frac{{\hbar ^2 }}{{2m^* }}\nabla ^2  + \left( {\tilde
J{\bf{n}}_r  \cdot {\bf{\sigma }} + V_0 d} \right)\delta \left( z
\right),
\end{equation}
where we assume a single spiral layer. $\tilde J =
\left\langle {J\left( z \right)} \right\rangle d $ refers to space
and momentum averages of the exchange coupling strength. It should be noted that the helimagnetic
field is sinusoidally space dependent. A multichannel-tunneling
picture should be considered and integer numbers of the helical wave
vector $\bar q$ could be absorbed or emitted in transmission and
reflection. With a plane wave incidence, the general wave function
of the incident, transmitted and reflected electrons can be written
as
\begin{equation}
\psi _{\texttt{NM}}^\sigma  \left( x,y,z \right) = e^{ik_x x} e^{ik_y y}
e^{ik_z z} \chi _\sigma   + e^{ik_y y} \sum\limits_{\sigma ',n}
{r_n^{\sigma \sigma '} e^{i {k_{x}^{n}  } x} e^{ - ik_z^n z} \chi
_{\sigma '} } ,
\end{equation}
\begin{equation}
\psi _{\texttt{FM}}^\sigma  \left( x,y,z \right) = e^{ik_y y}
\sum\limits_{\sigma ',n} {t_n^{\sigma \sigma '} e^{i {k_x^n} x}
e^{ik_z^{n\sigma '} z} \chi _{\sigma '} } ,
\end{equation}
with ${k_z} = \sqrt {2mE} \cos {\theta _{\texttt{in}}}/\hbar,$  ${k_x} = \sqrt {2mE} \sin {\theta _{\texttt{in}}}\cos {\phi _{\texttt{in}}}/\hbar,$ ${k_y} = \sqrt {2mE} \sin {\theta _{\texttt{in}}}\sin {\phi _{\texttt{in}}}/\hbar,$
$\theta_{\texttt{in}}$ and $\phi _{\texttt{in}}$ the incident polar and azimuthal angles, respectively. Here,
$n$ is an integer ranging from $ - \infty $ to $  \infty $ indexing
the diffraction order. And $
k_z^n  = \sqrt {2mE - \hbar ^2 k_y^2  - \hbar ^2 (k_x^n) ^2 }/\hbar $, $k_z^{n\sigma }  = \sqrt {2mE + \sigma \Delta  - \hbar ^2
k_y^2  - \hbar ^2 (k_x^n) ^2 }/ \hbar $, $k_x^n  = k_x  + n\bar q$.
By frame rotation, the FM eigenspinors can be obtained as
\begin{equation}
{\chi _ + } = \left( {\begin{array}{*{20}{c}}
{\cos \frac{\theta }{2}{e^{ - \frac{{i\phi }}{2}}}}\\
{\sin \frac{\theta }{2}{e^{\frac{{i\phi }}{2}}}}
\end{array}} \right),\begin{array}{*{20}{c}}
{}&{{\chi _ - } = \left( {\begin{array}{*{20}{c}}
{ - \sin \frac{\theta }{2}{e^{ - \frac{{i\phi }}{2}}}}\\
{\cos \frac{\theta }{2}{e^{\frac{{i\phi }}{2}}}}
\end{array}} \right),}
\end{array}
\label{eq2}
\end{equation}
corresponding to an electron spin parallel ($\sigma =+$) and
antiparallel ($\sigma =-$) to the magnetization direction in the FM
electrode, whose gauge is the same with eigenspinors
\begin{equation}
{\chi _ \uparrow } = \left( {\begin{array}{*{20}{c}}
1\\
0
\end{array}} \right),\begin{array}{*{20}{c}}
{}&{{\chi _ \downarrow } = \left( {\begin{array}{*{20}{c}}
0\\
1
\end{array}} \right),}
\end{array}
\label{eq3}
\end{equation}
of $\sigma _z$. Besides Eq. (\ref{eq2}), arbitrary gauge selection generates an arbitrary eigenspinor of ${\bf{m}}$ as
\begin{equation}
\chi _ + ^f = \left( {\begin{array}{*{20}{c}}
{\cos \frac{\theta }{2}{e^{ - \frac{{i\phi }}{2}}}}\\
{\sin \frac{\theta }{2}{e^{\frac{{i\phi }}{2}}}}
\end{array}} \right)f\left( \phi  \right),\begin{array}{*{20}{c}}
{}&{\chi _ - ^g = \left( {\begin{array}{*{20}{c}}
{ - \sin \frac{\theta }{2}{e^{ - \frac{{i\phi }}{2}}}}\\
{\cos \frac{\theta }{2}{e^{\frac{{i\phi }}{2}}}}
\end{array}} \right)g\left( \phi  \right),}
\end{array}
\label{eq5}
\end{equation}
with $f\left (\phi \right)$ and $g\left (\phi \right)$ arbitrary unitary complex functions (${\left| {f\left( \phi  \right)} \right|^2} = 1$ and ${\left| {g\left( \phi  \right)} \right|^2} = 1$).
 In the plane ($x$-$y$ plane) perpendicular to the
transport direction ($z$-axis), free motion of the electron is
assumed. Diffraction appears in the $x$-direction and $k_y$ is conserved under translational invariance.

The reflection ($r_n^{\sigma \sigma '}$) and transmission
($t_n^{\sigma \sigma '}$) amplitude in the $n$th diffraction order
can be numerically obtained from the continuity conditions\cite{Ref40} for $\Psi
(x,y,z)$ at $z=0$.
\begin{equation}
\Psi _{NM}^\sigma  \left( {x,y,0^ -  } \right) = \Psi _{FM}^\sigma
\left( {x,y,0^ +  } \right),
\end{equation}
\begin{equation}
\begin{array}{c}
 \frac{{\hbar ^2 }}{{2m_e }}\left. {\frac{{\partial \Psi _{NM}^\sigma  \left( {x,y,z} \right)}}{{\partial z}}} \right|_{z = 0^ -  }  + \left[ {V_0 d + {\bf{\tilde w}}\left( {\theta _r } \right)} \right]\Psi _{NM}^\sigma  \left( {x,y,0^ -  } \right) \\
  = \frac{{\hbar ^2 }}{{2m_e }}\left. {\frac{{\partial \Psi _{FM}^\sigma  \left( {x,y,z} \right)}}{{\partial z}}} \right|_{z = 0^ +  } , \\
 \end{array}
\end{equation}
with
\[
{\bf{\tilde w}}\left( {\theta _r } \right) = \tilde J\left[
{\begin{array}{*{20}c}
   {\cos \theta _r } & {\sin \theta _r }  \\
   {\sin \theta _r } & { - \cos \theta _r }  \\
\end{array}} \right].
\]
The continuity equation can be expressed in each
diffracted order. Transmissivity of a spin-$\sigma$ electron through the MF tunnel
junction with the incident wave vector $[k_x,k_y,k_z]$ to the $n$-th
diffracted order and spin-$\sigma '$ channel with the outgoing wave
vector $[k_x^n,ky,k_z^{n\sigma '}]$ reads
\begin{equation}
T_n^{\sigma \sigma '} \left( {E,k_y ,\theta _{xz} } \right) =
\frac{{\texttt{Re}\left( {k_z^{n\sigma '} } \right)  }}{{ k_z }}\left|
{t_n^{\sigma \sigma '} } \right|^2.
\end{equation}

The scattering matrix can be expressed as
\begin{equation}
\left( {\begin{array}{*{20}{c}}
{{b_{L \uparrow }}}\\
{{b_{L \downarrow }}}\\
{{b_{R \uparrow }}}\\
{{b_{R \downarrow }}}
\end{array}} \right) = {U^{fg \dag} }\left( {\begin{array}{*{20}{c}}
{{r_{ +  + }}}&{{r_{ -  + }}}&{t{'_{ +  + }}}&{t{'_{ -  + }}}\\
{{r_{ +  - }}}&{{r_{ -  - }}}&{t{'_{ +  - }}}&{t{'_{ -  - }}}\\
{{t_{ +  + }}}&{{t_{ -  + }}}&{r{'_{ +  + }}}&{r{'_{ -  + }}}\\
{{t_{ +  - }}}&{{t_{ -  - }}}&{r{'_{ +  - }}}&{r{'_{ -  - }}}
\end{array}} \right)U^{fg}\left( {\begin{array}{*{20}{c}}
{{a_{L \uparrow }}}\\
{{a_{L \downarrow }}}\\
{{a_{R \uparrow }}}\\
{{a_{R \uparrow }}}
\end{array}} \right),
\end{equation}
where the primed terms indicate inversive transport and $U^{fg}$ is the transform matrix from ${\sigma _z}$ to ${\bf{m}}$ representation in arbitrary gauge.
\begin{equation}
{U^{fg}} = \left( {\begin{array}{*{20}{c}}
1&0\\
0&1
\end{array}} \right) \otimes \left( {\begin{array}{*{20}{c}}
{\cos \frac{\theta }{2}{e^{ - i\frac{\phi }{2}}}f}&{\sin \frac{\theta }{2}{e^{i\frac{\phi }{2}}}f}\\
{ - \sin \frac{\theta }{2}{e^{ - i\frac{\phi }{2}}}g}&{\cos \frac{\theta }{2}{e^{i\frac{\phi }{2}}}g}
\end{array}} \right).
\end{equation}
 We consider spin pumping driven by ferromagnet magnetization precession (see Fig. 1), so this transformation is necessary. The time-dependent parameter is $\phi$, which is both a physical quantity and also a gauge factor in $f$ and $g$. Transport of different diffraction order would not be correlated by time-dependent variation. Therefore, the adiabatically pumped $2 \times 2$ tensor current in the NM electrode can be calculated by\cite{Ref3, Ref22}
 \begin{equation}
 {\hat I} = \frac{{e\omega }}{{4{\pi ^2}}}\int_0^{2\pi } {dXk_{\max }^2\sin {\theta _{\texttt{in}}}d{\theta _{\texttt{in}}}d{\phi _{\texttt{in}}}\sum\limits_{m =  - \infty }^\infty  {{\mathop{\rm Im}\nolimits} {{\left[ {\hat S_m^\dag \frac{{\partial {{\hat S}_m}}}{{\partial X}}} \right]}_{LL}}} } ,
 \end{equation}
where ${k_{\max }} = \sqrt {2mE} /\hbar $ and $m$ is the diffraction order.
The pumped charge and spin current follows as
 \begin{equation}
 \begin{array}{l}
{I_c} = {{\hat I}_{11}} + {{\hat I}_{22}},\\
I_s^z = {{\hat I}_{22}} - {{\hat I}_{11}},\\
I_s^x =  - \left( {{{\hat I}_{12}} + {{\hat I}_{21}}} \right),\\
I_s^y =  - i\left( {{{\hat I}_{12}} - {{\hat I}_{21}}} \right),
\end{array}
\end{equation}
with the spin angular momentum flow defined in $\hbar {\bf{\sigma }} \cdot {\bf{I}}/2$.

\section{Numerical results and interpretations}

We consider spin pumping driven by magnetization precession in the NM/MF/FM heterostructure (see Fig. 1). In numerical calculations, the NM Fermi energy
$E_F$ is chosen to be $5.5$ eV. Different values of $E_F$ would not change the primary pumping properties. The spatial average of
the helimagnetic exchange coupling strength $\tilde J = 0.2$
${\rm{eV}} \cdot {\rm{nm}}$, which is reasonable compared to the
Fermi energy. Periods of short-period and long-period helimagnets are $3$-$6$ nm and $18$-$90$ nm, respectively\cite{Ref44}. In our model
we set the period to be $10$ nm and hence $\bar q=2 \pi /10$
$\texttt{nm} ^{-1}$. The magnetization of the FM electrode precesses anticlockwise around the $z$-axis. Zeeman splitting in the
FM electrode $\Delta =2$ eV. Barrier height of the MF oxide plane
$V_0 =0.5$ eV and width $d=2$ nm. We consider diffraction orders of $0$ and $\pm 1$ and numerically proved that keeping the three orders is sufficient for physical $\bar q$'s as higher orders decrease exponentially. The magnitude of the pumped current is in the order of $10^{-17}$ A. So it is far below the strength to rotate the helimagnet spiral or induce Gilbert damping in the FM electrode.

During transmission, the incident electron with wave
vector $[k_x,k_z]$ would absorb or emit $n \bar q$ from the
helimagnet and be diffracted into tunnels with wave vector
$[k_x^n,k_z^{n \sigma}]$.
Numerical results of the transmission of a single incident beam with $\theta _{\texttt{in}}$ and $\phi _{\texttt{in}}$ fixed are shown
in Figs. 1 and 2. It can be seen that zero-order spectrum governs the spin-conserved
transmission whereas first-order diffracted spectrum governs the spin-flipped transmission due to the grating effect in the spin space. The second-order spectrum is four orders smaller, which justifies the $0$, $\pm 1$ order cutoff. The exponential decrease in diffracted orders agrees with general grating properties. The $0$, $\pm 1$ order cutoff was also justified by numerical confirmation of the unitarity of the scattering matrix $\hat S$ including the $0$, $\pm 1$ orders. The transmission of one-way
incident light through sinusoidal gratings is delta-function-like
strict lines. Analogously, direction of transmission of one-way
incident electron through sinusoidal helimagnet is discrete strict
lines of different grating orders and the spin is conserved or
flipped. For arbitrary $\textbf{m}$ relative
to the chirality of the helimagnet, $+ 1$ and $-1$ order
transmission may not be symmetric. Physically, an electron with spin
polarization along the FM magnetization is transmitted in different
direction with its spin rotating an angle.

It can be seen from Fig. 2 that the $1$ and $-1$ order diffracted transmission varies with $\phi $ in trigonometric functions. The $0$ order transmission varies with $\phi $ much more slowly than the diffracted $\pm 1$ orders, which can be clearly seen from the scale extension in Fig. 3. Also the $0$ order transmission varies with $\phi $ in trigonometric functions with higher harmonics and the variation range is one to two digits smaller than the diffracted $\pm 1$ orders. These observations give rise to the pumping properties driven by cyclic modulation of $\phi$, which is shown in Fig. 4. Two prominent properties highlight the pumping physics in the NM/MF/FM heterostructure. The first is the non-vanishing pumped charge current. It is well known that in FM based tunnel junctions, nonzero spin current with zero charge current can be generated by cyclic magnetization precession. In the latter structure, spatial uniformity gives rise to exactly the same charge scattering matrix for different magnetization azimuthal angles. And transmission difference in the spin space accumulates during precession. However the spatial symmetry is broken by the helimagnet spiral. Non-zero spin as well as non-zero charge current is driven out by ferromagnet precession.

The other prominent property is gauge dependence of the pumped charge and spin currents. The considered situation is special in two points. The first is spatial nonuniformity. As a result, spin-dependent diffraction occurs. And also the diffraction effect cannot be averaged out by the cyclic integral of $\phi$. The second one is that the time-dependent parameter $\phi$ is physically the precession angle and nonphysically the gauge factor simultaneously. The two roles play independently. This point makes distinct demonstration at the two poles of the precession sphere. Fig. 4 shows the pumped charge and spin current using the eigenspinors of Eq. (\ref{eq2}) and Fig. 5 shows its order expansion of one incident electron beam. It can be seen that the pumped current does not vanish at the two poles of the precession sphere. We have calculated the transmission spectrum at the two poles and find that all of the three orders of the transmission probabilities do not vary with the magnetization azimuthal angle $\phi$ at the two poles and and the nonzero pumped current is a pure result of the phase of the transmission, the latter of which is not unusual in quantum pumps. Also we replace the MF helimagnet layer by a plane delta barrier and reobtained the magnetization-precession-driven pumped spin current of the FM tunnel junction and it vanishes at the two poles of the precession sphere. From Fig. 5 it can be seen that the nonzero pumped current at the two poles is a pure effect of diffraction with the zero-order pumped vanishing at the two poles.

Although the numerical results and analysis given above seem self-consistent, it is nonphysical. Precession at zero precession angle makes no sense and to say the least the area of the cyclic loop vanishes. Eq. (\ref{eq2}) is the eigenspinor obtained by frame rotation. It should have the same gauge phase as the eigenspinor (\ref{eq3}) of ${\sigma _z}$. However, if we make $\theta =0$ in Eq. (\ref{eq2}), it could not return to Eq. (\ref{eq3}). The remaining $\phi$ is a pure gauge phase but over the cyclic loop integral its effect accumulates instead of cancelling out giving rise to nonvanishing pumped current at zero $\theta$. The problem could be solved by the gauge transformation Eq. (\ref{eq5}) with
\begin{equation}
f\left( \phi  \right) = {e^{\frac{{i\phi }}{2}}},\begin{array}{*{20}{c}}
{}&{g\left( \phi  \right) = {e^{ - \frac{{i\phi }}{2}}}.}
\end{array}
\label{eq4}
\end{equation}
Numerical results of the pumped current in this gauge are shown in Fig. 6. It can be seen that the pumped current vanishes at the $\theta =0$ pole. It does not vanish at the $\theta = \pi$ pole since the gauge of Eq. (\ref{eq4}) also could not return to the eigenspinors of $-\sigma _z$, which are the swap of the two spinors of Eq. (\ref{eq3}).
Also under gauge transformation Eq. (\ref{eq5}) with
\begin{equation}
f\left( \phi  \right) = {e^{-\frac{{i\phi }}{2}}},\begin{array}{*{20}{c}}
{}&{g\left( \phi  \right) = {-e^{  \frac{{i\phi }}{2}}},}
\end{array}
\label{eq6}
\end{equation}
the pumped current vanishes at the $\theta =\pi$ pole and is nonzero at the $\theta=0$ pole, which is not shown in the manuscript to avoid tediousness.
In consideration of the pumping properties, gauges (\ref{eq5}) and (\ref{eq2}) are equivalent only when $f(\phi)=g(\phi)$. In the two cases the relation is not satisfied.

 The gauge selection in eigenspinors seldom matters. But it really does sometimes. Therefore the numerical results raised a question in the considered situation: which gauge is appropriate and why? We would like to argue that the question is nontrivial in at least two aspects. The first is that gauge difference in the pumped current only occurs in the contribution from diffracted orders. We show the order expansion of the pumped current of gauge Eq. (\ref{eq4}) in Fig. 7. It can be seen that the zero order contribution of the pumped current is both qualitatively and quantitatively identical to that of the rotation-frame gauge in Fig. 5. Obtained numerical results of other gauges show the same properties. This is probably the reason why the gauge difference is not encountered in spatially uniform quantum pumps without diffraction. The second is that physically sound results at the two poles of the precession sphere can only be achieved by two un-equivalent gauges. Usually in situations when the intrinsic phase of eigenspinors matters frame rotation gauge of Eq. (\ref{eq2}) is thought to secure correctness. However here it leads to nonphysical results at the two poles. From Figs. 4 to 7, it can be seen that diffraction induced gauge dependence of the pumped current is spectacular, which could not be explained by existed theory to our knowledge.

\section{Conclusions}

In this work, we considered the adiabatically pumped charge and spin current driven by precessing ferromagnet in the NM/MF-helimagnet/FM heterostructure. In this structure, space modulation of the helimagnet spin spiral gives rise to diffraction in the transmission spectrum. It is found that the usually neglected gauge difference in the ferromagnet eigenspinor cannot be overlooked trivially. Usually ubiquitous gauge phase obtained by frame rotation of the $\sigma _z$ eigenspinors gives non-vanishing pumped current at the two poles of the precessing sphere, which makes no sense. We numerically confirmed the unitarity of the scattering matrix of the combined zero and $\pm 1$ order diffracted states, which justifies the three order cutoff. We also show that the gauge dependence is a pure effect of diffraction with the zero order contribution gauge independent. By selecting two distinctive gauges at the two poles, physically sound vanishing pumped current can be obtained separately. These results raised the unsettled question of a self-consistent gauge definition in the considered situation.

\section{Acknowledgements}

The author acknowledges enlightening discussions with Jamal Berakdar, Wen-Ji Deng, Zhi-Lin Hou, and Wei-Kang Fan. This project was supported by the National Natural Science
Foundation of China (No. 11004063) and the Fundamental Research
Funds for the Central Universities, SCUT (No. 2014ZG0044).

\clearpage

\begin{figure}[h]
\includegraphics[height=14cm, width=16cm]{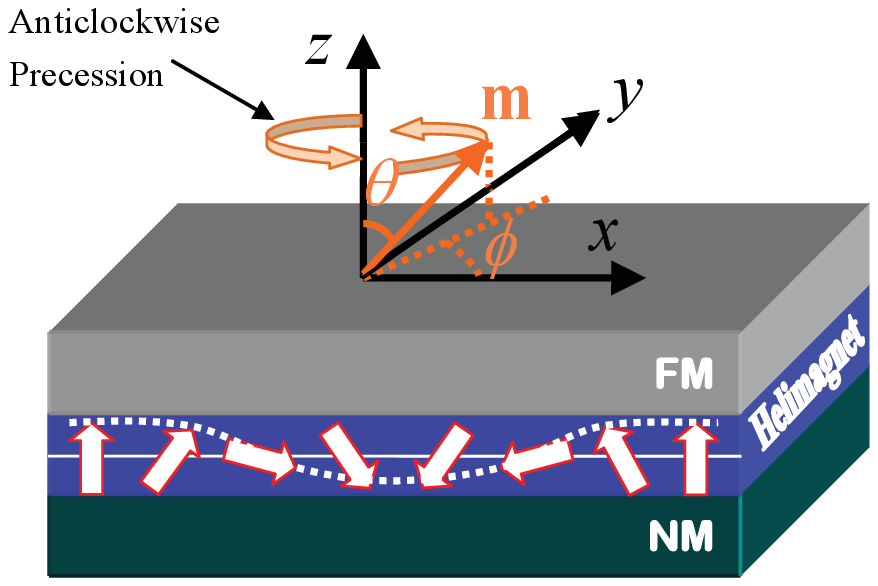}
\caption{Schematics of the spin pump based on a normal-metal/multiferroic-helimagnet/ferromagnet triple-layer heterostructure. In the helimagnet, the spin spirals in the $x$-$z$ plane in the trigonometric function. The white arrow and dotted line indicate the spin and the spiral envelope function respectively. The thin orange arrow indicates the direction of the ferromagnetic magnetization $\bf{m}$ with polar angle $\theta $ and azimuthal angle $\phi $. The wide orange arrow indicates the anticlockwise magnetization precession around $z$-axis.}
\end{figure}

\clearpage

\begin{figure}[h]
\includegraphics[height=10cm, width=14cm]{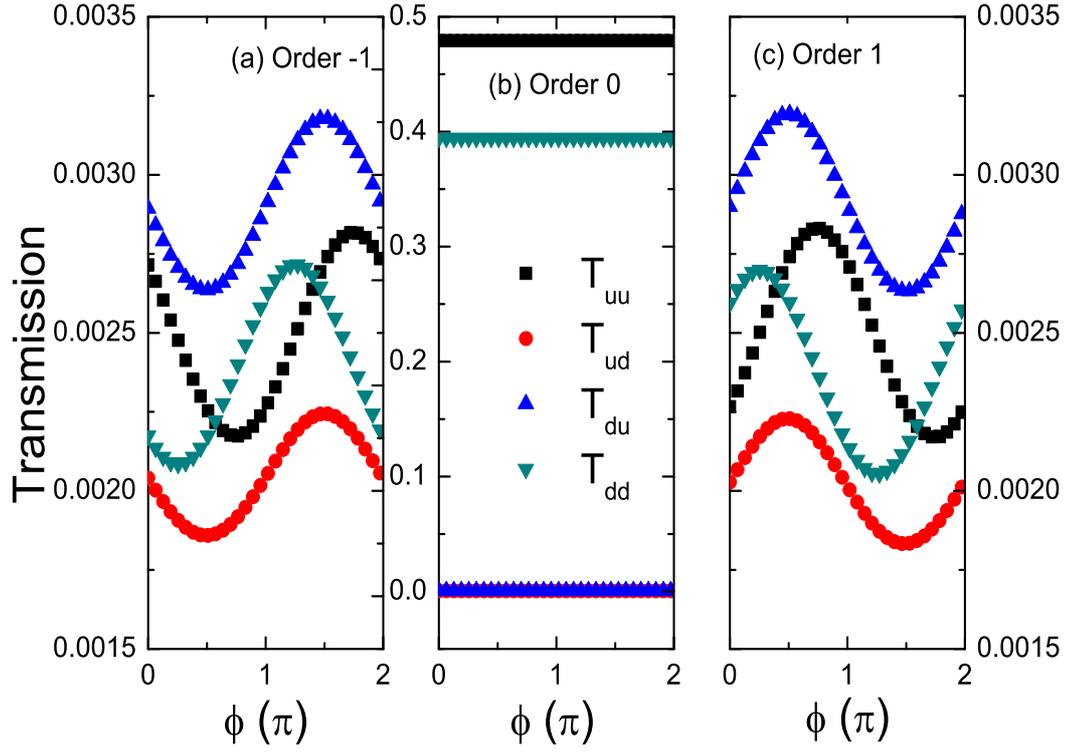}
\caption{Spin-dependent transmission of different diffracted orders. $\theta$=0.5, $\theta_{\texttt{in}}=0.2$, $\phi_{\texttt{in}}=0.5$ in radian. }
\end{figure}

\begin{figure}[h]
\includegraphics[height=10cm, width=14cm]{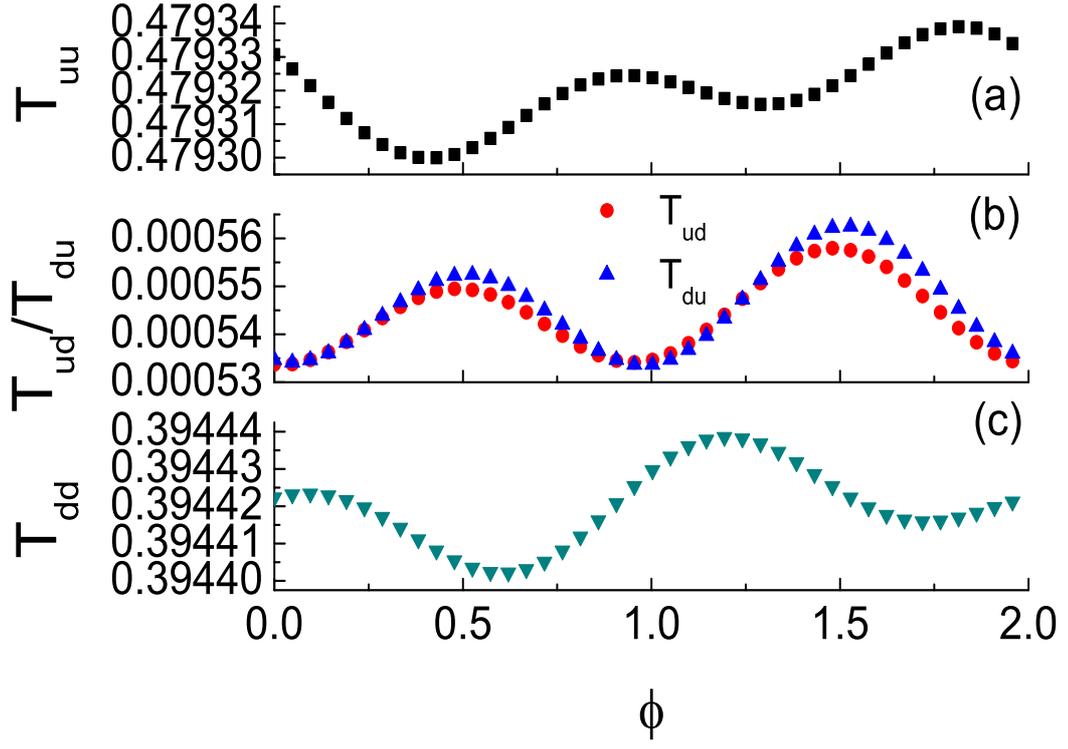}
\caption{Zero-order spin-dependent transmission. $\theta$=0.5, $\theta_{\texttt{in}}=0.5$, $\phi_{\texttt{in}}=0.2$ in radian. }
\end{figure}

\begin{figure}[h]
\includegraphics[height=10cm, width=14cm]{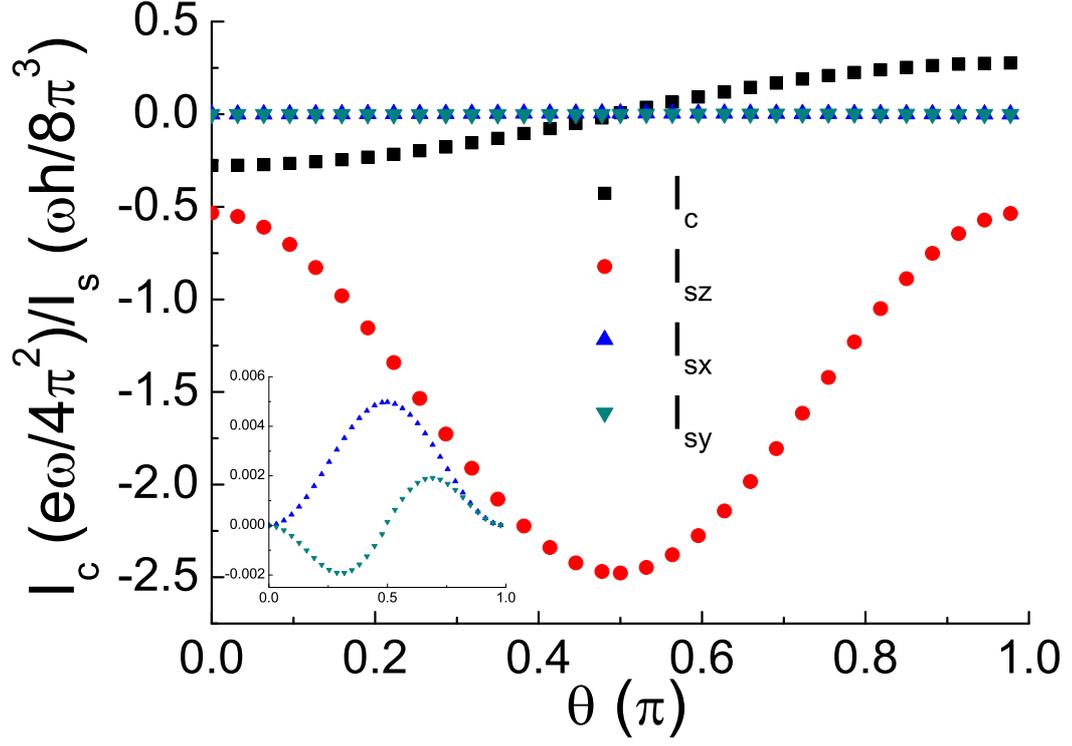}
\caption{Pumped charge and spin currents as a function of the precession angle in the gauge of Eq. (\ref{eq2}). Inset is the zoom-in of the pumped spin angular momentum flow in the $x$ and $y$ directions.}
\end{figure}

\begin{figure}[h]
\includegraphics[height=10cm, width=14cm]{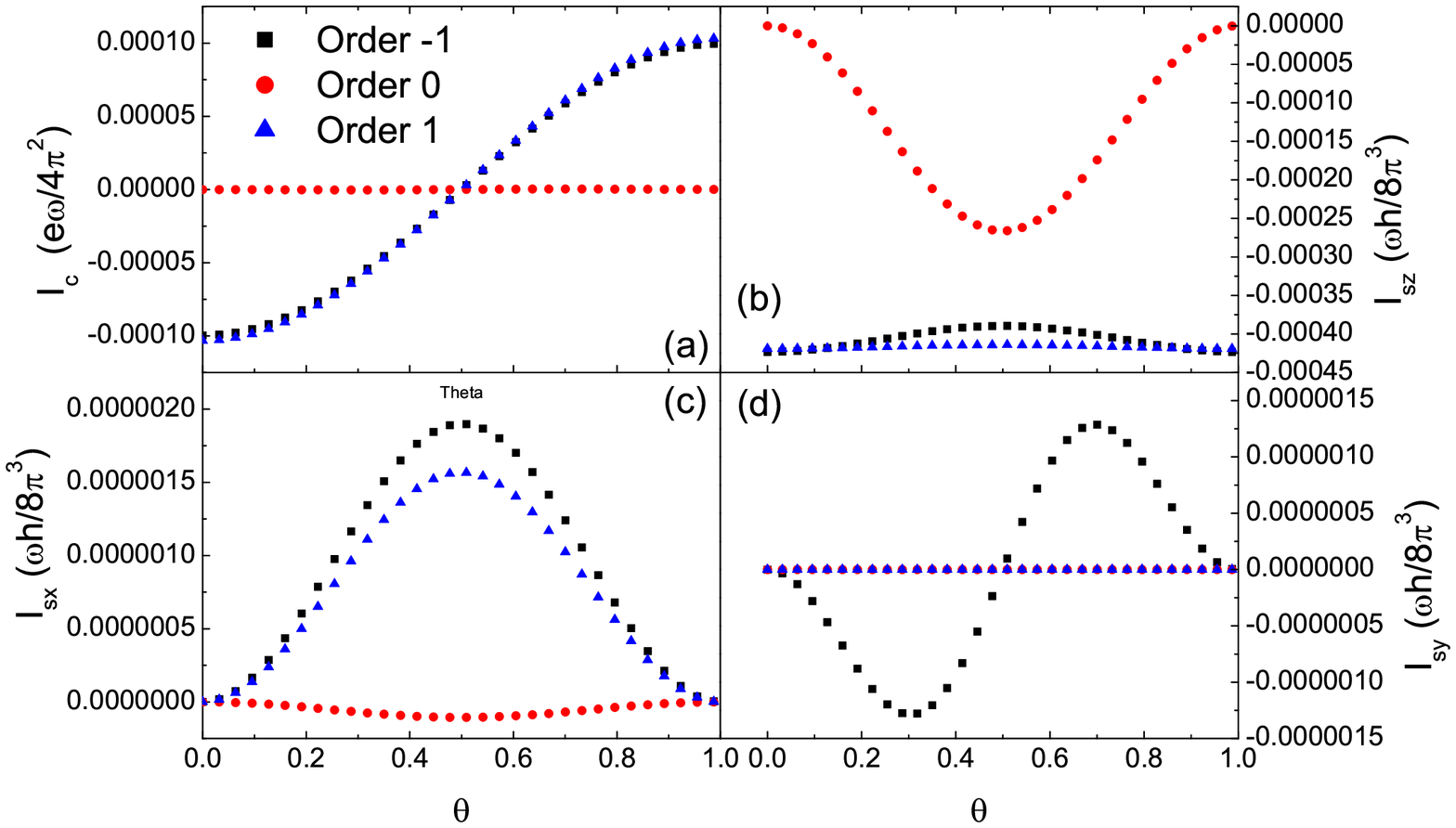}
\caption{Order expansion of the pumped charge and spin currents as a function of the precession angle in the gauge of Eq. (\ref{eq2}). Shown is results of a single incident beam with $\theta_{\texttt{in}}=0.5$ and $\phi_{\texttt{in}}=0.2$ in radian.}
\end{figure}

\begin{figure}[h]
\includegraphics[height=10cm, width=14cm]{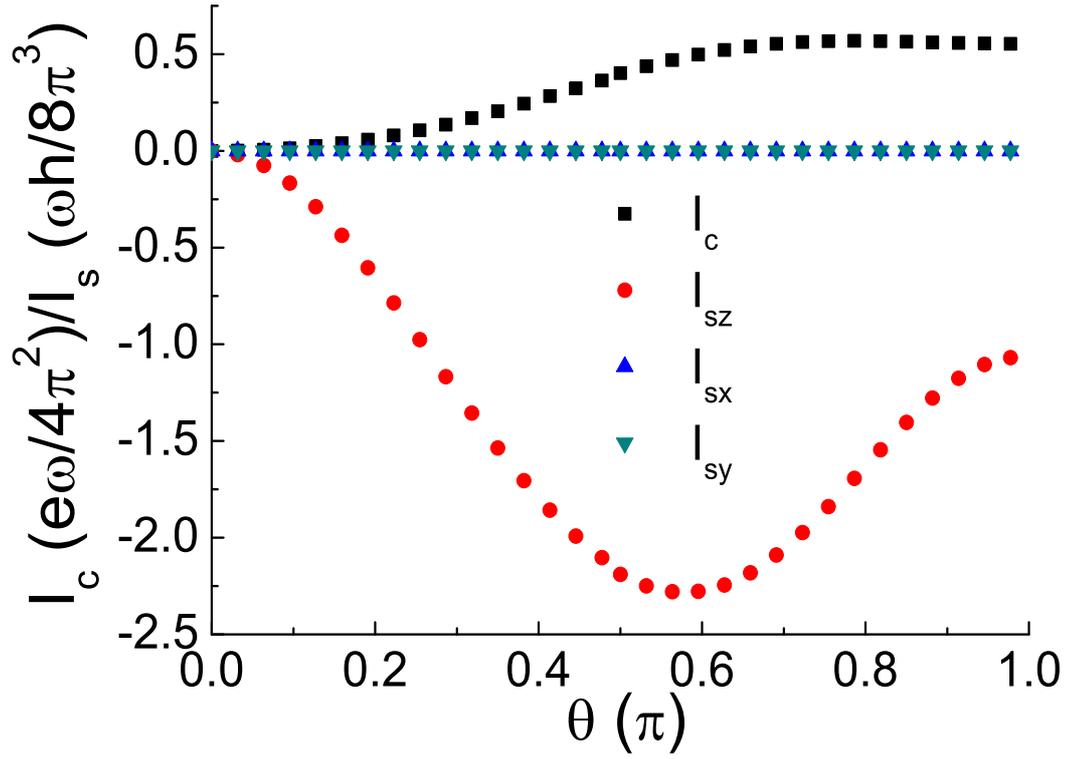}
\caption{Pumped charge and spin currents as a function of the precession angle in the gauge of Eq. (\ref{eq4}). $I_{sx}$ and $I_{sy}$ are zero to the accuracy of $10^(-10)$ order.}
\end{figure}

\begin{figure}[h]
\includegraphics[height=10cm, width=14cm]{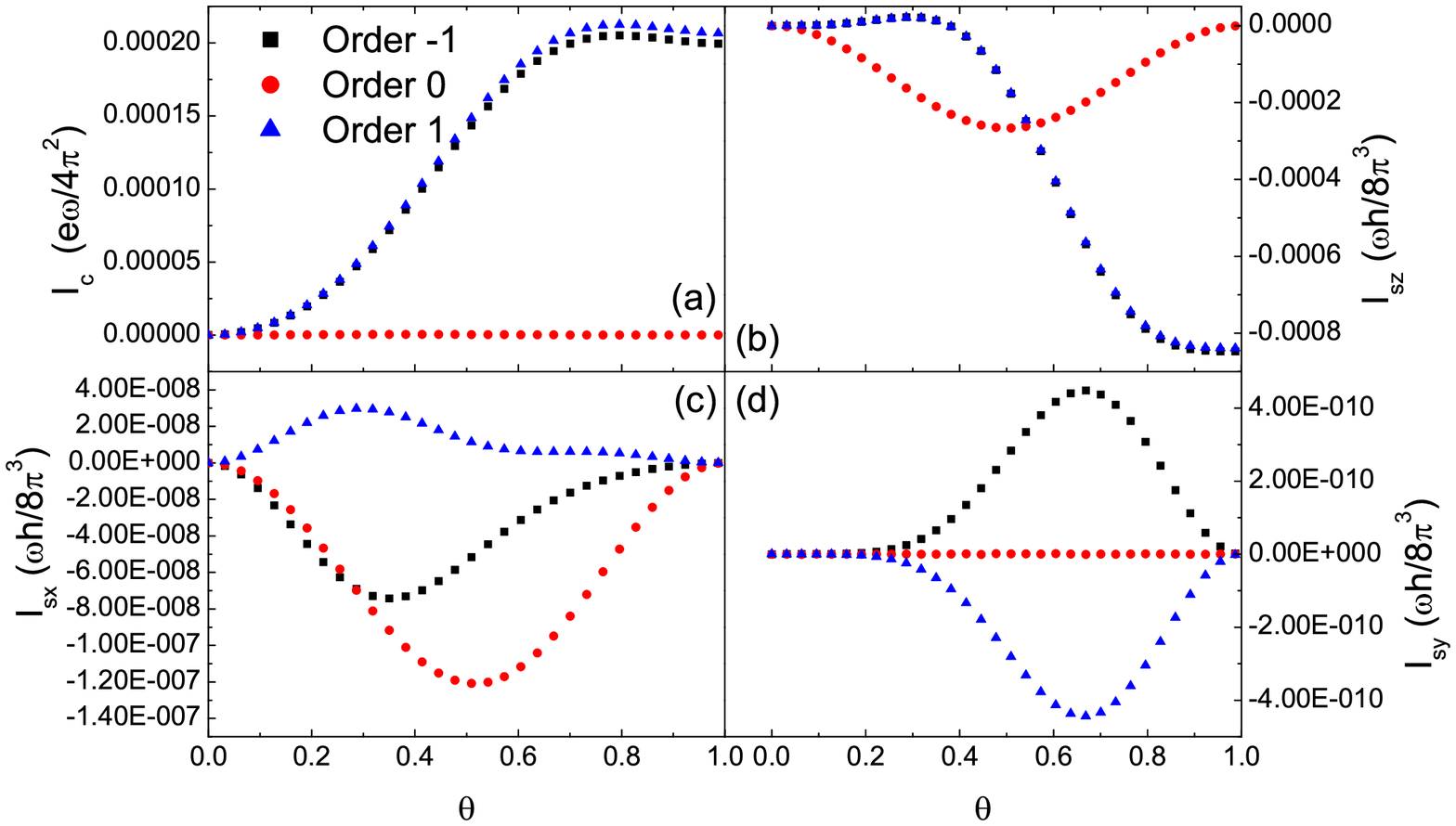}
\caption{Order expansion of the pumped charge and spin currents as a function of the precession angle in the gauge of Eq. (\ref{eq4}). Shown is results of a single incident beam with $\theta_{\texttt{in}}=0.5$ and $\phi_{\texttt{in}}=0.2$ in radian. }
\end{figure}

\clearpage

\clearpage

\end{document}